\documentclass[10pt]{article}
 \usepackage{latexsym,graphicx}
 \usepackage[left=2.5cm,top=2.5cm,right=2.5cm,bottom=1.5cm]{geometry}
 \usepackage{textcomp}
 \usepackage{amsmath}
 \usepackage{amssymb}
 \usepackage{graphicx}
 \begin{document}
 \begin{center}
\Large{\bf{ FLRW Accelerating Universe with Interactive Dark Energy }}

\vspace{2mm} \vspace{2mm}
\normalsize{G. K. Goswami$^1$, Anirudh Pradhan$^2$, A. Beesham$^3$ }\\
\vspace{2mm}
\normalsize{$^1$ Department of Mathematics, Kalyan P. G. College, Bhilai-490 006, C. G., India}\\
\vspace{2mm}
\normalsize{$^1$ Email: gk.goswami9@gmail.com} \\
\vspace{2mm} \normalsize{$^2$ Department of Mathematics, Institute of Applied
Sciences and Humanities, G L A University,
    Mathura-281 406, Uttar Pradesh, India  \\
    \vspace{2mm}
    E-mail: pradhan.anirudh@gmail.com} \\
\vspace{2mm}
\normalsize{$^3$ Department of Mathematical Sciences, University of Zululand, Kwa-Dlangezwa 3886, South Africa  \\
    \vspace{2mm}
    E-mail: beeshama@unizulu.ac.za}
\end{center}
\vspace{10mm}
 \begin{abstract}
    We have developed an accelerating  cosmological model for the present  universe which  is phantom for the period $ (0 \leq z \leq 1.99)$
     and quintessence phase for $(1.99 \leq z \leq 2.0315)$. The universe is assumed to be filled with barotropic and dark energy(DE) perfect
      fluid in which DE interact with matter. For  a  deceleration parameter(DP) having decelerating-accelerating transition phase of  universe,
      we assume hybrid expansion law for scale factor. The transition red shift for the model is obtained as $z_t = 0.956$.
       The model  satisfies current observational constraints.
 \end{abstract}
\section{Introduction}
The cosmological principle (CP), which states that  there is no privileged position in the universe and it is as such spatially homogeneous
and isotropic, is the backbone of any cosmological model of the universe.  Friedman-Lemaitre-Robertson-Walker (FLRW) line element
fits best with the CP. The FLRW model, in the background of a perfect fluid distribution of matter, represents an expanding and
decelerating universe. However the latest findings on observational grounds during the last three decades by various cosmological
missions \cite{ref1}$-$\cite{ref17} all confirm  that our universe is undergoing an accelerating expansion. In $\Lambda$CDM cosmology
 \cite{ref18, ref19}, the  $\Lambda$- term is used as a candidate of DE with equation of state
 $ p_ \Lambda = \rho _ \Lambda = \frac{-\Lambda c^4}{8\pi G}$. However, the model suffers from, inter alia,
fine tuning and cosmic coincidence problems \cite{ref20}. Any acceptable cosmological model must explain the  accelerating  universe.\\

 Of late,  many authors  \cite{ref21}$-$\cite{ref26}  presented  DE models in which the  DE is considered in a conventional manner
as a fluid with an EoS parameter $\omega_{de} = \frac{p_{de}}{\rho_{de}}$. It is
assumed that our universe is filled with two types of perfect fluids in which one is
a barotropic fluid (BF) which has positive pressure and creates  deceleration in
the universe. The other one is a DE fluid which has negative pressure and
creates acceleration in the universe. Both fluids have different EoS parameters.
Zhang and Liu \cite{ref27} have constructed DE models with higher derivative
terms. Liang {\it et al.} \cite{ref28} have investigated two-fluid dialation model of
DE. The modified Chaplygin gas with interaction between holographic DE and
dark matter has been discussed by
Wang {\it et al.} \cite{ref29}.\\

 Recently, it has been discovered that the interaction between DE and dark matter(DM) offers an attractive alternative to the
standard model of the cosmology \cite{ref30,ref31}. In these works the motivation to study interacting DE model arises from
high energy physics. In recent work Risalti and Lusso \cite{ref32} and Riess {\it et al.} \cite{ref33} stated that a rigid $\Lambda$
is ruled out by $~4\sigma$ and allowing for running vacuum favored phantom type
DE ($\omega < -1$) and $\Lambda$ CDM is claimed to be ruled out by $4.4\sigma$ motivating the study of interactive DE models. Interacting DE
models \cite{ref34}$-$\cite{ref38} lead to the idea that DE and DM do not evolve separately but interact with each other non gravitationally
(see recent review \cite{ref39} and references there in.).\\

In this paper,  we have developed an accelerating  cosmological model for the present  universe which  is phantom for the period
 $ (0 \leq z \leq 1.99)$ and quintessence phase for $(1.99 \leq z \leq 2.0315)$. The universe is assumed to be filled with barotropic
  and dark energy(DE) perfect fluid in which DE interact with matter.  For  a  deceleration parameter(DP) having decelerating-accelerating
   transition phase of  universe, we assume hybrid expansion law for scale factor. The transition red shift for the model is obtained as $z_t = 0.956$.
    The model  satisfies current observational constraints.

\section{Basic field equations}

The dynamics of the universe is governed by the Einstein's field equations (EFEs) given by

\begin{equation}
\label{eq1}
R_{ij}-\frac{1}{2}Rg_{ij} =-\frac{8\pi G}{c^{4}}T_{ij},
\end{equation}
where $R_{ij}$ is the Ricci tensor, $R$ is the scalar curvature, and $T_{ij}$ is the
stress-energy tensor taken as $T_{ij} = T_{ij}(m)+T_{ij}(de).$ We assume that our
universe is filled with two types of perfect fluids (since homogeneity and isotropy
imply that there is no bulk energy transport), namely  baryonic fluid and dark
energy. The energy-momentum tensors of the contents of the universe are
presented as follows: ( The subscripts $m$ and $de$ denote ordinary matter and
dark energy, respectively.) $T_{ij}(m)=\left(\rho_m + p_m\right)u_{i}u_{j}-p_m
g_{ij}$ and $T_{ij}(de)=\left(\rho_{de}+p_{de}\right)u_{i}u_{j}-p_{de} g_{ij}$. In
standard spherical coordinates
 $x^{i} = (t, r, \theta, \phi)$, a spatially homogeneous and isotropic FLRW
 line-element is the following (in units $c = 1$)
\begin{equation}
\label{eq2}
ds{}^{2}=dt{}^{2}-a(t){}^{2}\left[\frac{dr{}^{2}}{(1+kr^{2})}+r^{2}({d\theta{}^{2}+sin{}^{2}\theta
    d\phi{}^{2}})\right],
\end{equation}
where (i)  k=-1 is closed universe (ii)  k=1 is open universe and (iii) k=0 is
spatially flat universe. Solving EFEs (\ref{eq1}) for  the FRW metric (\ref{eq2}), we
get the following equations of dynamic cosmology.
\begin{equation}
\label{eq3}
2\frac{\ddot{a}}{a}+H^{2} = -8\pi G p + \frac{k}{a^{2}}
\end{equation}
and
\begin{equation}
\label{eq4}
H^{2} =  \frac{8\pi G}{3}\rho + \frac{k}{a^{2}},
\end{equation}
where $H=\frac{\dot{a}}{a}$ is the Hubble constant. Here an over dot means differentiation with respect to cosmological
time $t$. We have deliberately put the curvature term on the right of Eqs. (\ref{eq3}) and (\ref{eq4}), as this term is made
to acts like an energy  term. For this, we assume that  the density and pressure for the curvature energy are as follows
$\rho_{k}=\frac{3 k}{8\pi Ga^{2}},  p_{k}=-\frac{k}{8\pi Ga^{2}}.$ With this choice, Eqs. (\ref{eq3}) and (\ref{eq4}) are read as
\begin{equation}
\label{eq5}
2\frac{\ddot{a}}{a}+H^{2} = -8\pi G (p+p_{k})
\end{equation}
and
\begin{equation}
\label{eq6}
H^{2}=\frac{8\pi G}{3}\,(\rho+\rho_{k}).
\end{equation}
The energy density $\rho$ in Eq. (\ref{eq6}) is comprised of two types of energy,
namely matter and dark energy $\rho_m$ and $\rho_{de}$, where as the
pressure `$p$' in  Eq. (\ref{eq5})  is  comprised of pressure due to matter and
pressure due to dark energy. We can express $\rho = \rho_m + \rho_{de}$ and
$p =p_m+p_{de}.$

\section{ Energy conservation laws \& densities}
The energy conservation law[ECL]~ $T^{ij}_{;j}=0$~ provides the following well known equation amongst the density $\rho$, pressure $p$
and Hubble constant $H$, $\dot{\rho}+3H(p+\rho)=0$,
where  $\rho=\rho_{m}+\rho_{de}+\rho_{k}$
and
$p=p_{m}+p_{de}+p_{k},$
are the total density and pressure of the universe, respectively.
We see that $\rho_{k}$ and $p_{k}$ satisfy ECL independently,
i.e.$\dot{\rho_{k}}+3H(p_{k}+\rho_{k})=0$, so  that
$\frac{d}{dt}{(\rho_{m}+\rho_{de})}+3H(p_{m}+p_{de}+\rho_{m}+\rho_{de})=0$.
We assume that DE interacts with and transforms energy to baryonic matter.
For this, the continuity equations for the dark  and baryonic fluids can be written as follows
\begin{equation}
\label{eq7}
\dot{\rho_{m}}+3H(p_{m}+\rho_{m})= Q
\end{equation}
and
\begin{equation}
\label{eq8}
\dot{\rho_{de}}+3H(p_{de}+\rho_{de})= -Q
\end{equation}

The quantity Q represents the  energy transfer from DE to baryonic matter, so
we take  $ Q \ge 0$. We follow Amendola {\it el al.} \cite{ref40} and Gou {\it el
al.} \cite{ref41}, to assume that
\begin{equation}\label{eq9}
Q = 3 H \sigma \rho_{m},
\end{equation}
where $\sigma$ is a coupling constant and is positive.

At present our universe is dust filled, so we take $p_{m}=0$. Integrating Eqs.
(\ref{eq7}) and  (\ref{eq8}) with the help of Eq. (\ref{eq9}), we get $\rho_{m} =
(\rho_{m}) _0\left (1+z\right)^{3(1-\sigma)}$ and $\rho_{de} = (\rho_{de}) _0~ exp
\left( 3\int^z_0 \frac{(1+\omega_{de})dz}{1+z}\right)$, where we have put
$\frac{a_0}{a}=1+z$. Clearly DE helps in the expansion of the universe through
energy transfer. The EoS for the curvature energy is obtained as
$p_{k}=\omega_{k}\rho_{k}~\mbox{where}~ \omega_{k}=-1/3.$ This gives
$\rho_{k} = (\rho_{k})_{0} (1+z)^2.$ The critical density and   density parameters
for energy density, dark energy and curvature density  are, respectively, defined
by $\rho_{c}=\frac{3H^{2}}{8\pi G}, \Omega_{m}=\frac{\rho_{m}}{\rho_{c}},
 \Omega_{de}=\frac{\rho_{de}}{\rho_{c}}$ and $\Omega_{k}=\frac{\rho_{k}}{\rho_{c}}$, where $\rho_{c}$, $\Omega_{m}$, $\Omega_{de}$
  and $\Omega_{k}$ are the critical density, matter energy ,
dark energy  and curvature energy parameters  respectively. \\

With these in hand, we can write the FRW field equations as follows
\begin{equation}
\label{eq10}
H^2=H^{2}_{0}\left[(\Omega_{m})_{0} \left(1+z\right)^{3(1-\sigma)}+(\Omega_{k})_{0}
\left(1+z\right)^{2} \right] + H^2 \Omega_{de},
\end{equation}
and
\begin{equation}
\label{eq11}
2q = 1 - \frac{H^2_0}{H^2}
 (\Omega_{k})_{0} \left(\frac{a_0}{a}\right)^2+ 3\omega_{de}\Omega_{de} ,
\end{equation}
where $q$ is DP defined by $q=-\frac{\ddot{a}}{aH^2}.$
The purpose of this paper is to investigate the evolution of   $\omega_{de}$ over red shift or time and to match it with the observational constraint.
\section{ Hybrid Scale Factor with Plank Results }

We have only two equations and the scale factor `$a$', pressure $p$ and energy density $\rho$ to be determined. So we have to
use a certain ansatz. As motivation for the ansatz, we note some important solutions. The De Sitter universe
has scale factor $a(t)= exp(\Lambda t)$  where $\Lambda$ is the positive cosmological. Later on, FRW cosmological models were
proposed in which Einstein and De Sitter gave the power law expansion law $a(t)= t^{2/3}$ for flat space-time.
Off late, during the last three decades, researchers are working with accelerating expanding models describe a transition
from deceleration to acceleration. \\

In the literature a constant deceleration parameter \cite{ref42}$-$\cite{ref45} and references therein, has been used to give
a power or exponential law. As it has been discussed in the introduction that in view of the recent observations of Type Ia
supernova \cite{ref1}$-$\cite{ref5}, WMAP collaboration \cite{ref12,ref46,ref47}, and Planck Collaboration \cite{ref17}
there is a need of a time-dependent deceleration parameter which describe decelerated expansion in the past and accelerating
expansion at present, so there must be a transition from deceleration to acceleration. The deceleration parameter must
show the change in signature \cite{ref48}$-$\cite{ref50}.\\

Now, we consider a well-motivated ansatz considered by Abdussattar and Prajapati \cite{ref51}, which puts a constraint
on the functional form of the deceleration parameter $q$ as
\begin{equation}
\label{eq12}
q=\frac{k n}{(k+ t)^{2}}-1,
\end{equation}
where $k>0$ (dimension of square of time) and $ n> 0$ (dimensionless) are constants.
For such choice of the scale factor, we see that  $q=0$ when $t=\sqrt{k n}-k$. We get $q>0$ (i.e. decelerated expansion)
for $t<\sqrt{k n}-k$ and $q<0$ (i.e. accelerated expansion) for $t>\sqrt{k n}-k$.
Integrating $q=-\frac{\ddot{a}a}{\dot{a}^2}$, we find the scale factor as
\begin{equation}
\label{eq13}
a(t)=c_2 \exp\int\frac{dt}{\int (1+q)dt+c_1},
\end{equation}
where $c_1, c_2$ are integrating constants.\\

Choosing appropriate values of the constants ($c_{1}=n$ and $c_{2}=1$), one can integrate Eq. (\ref{eq13}) with the
help Eq. (\ref{eq12}) to get the scale factor as
\begin{equation}
\label{eq14}
a(t)=t^{\alpha} \exp(\beta t ),
\end{equation}
where $\alpha > 0$ and $\beta > 0$ are constants.\\

    Akarsu et al. \cite{ref52} also used hybrid expansion law [HEL] with scalar field reconstruction of observational constraints
    and cosmic history. Avil$e'$s et al. \cite{ref53} used HEL with integrating cosmic fluid. Several authors  \cite{ref54}$-$\cite{ref61}
    have considered the HEL for solving different cosmological problems in general relativity and $f(R,T)$ gravity theories. Some work is
    done by Moraes \cite{ref62} and  Moraes et al. \cite{ref63}. Recently, Moraes and Sahoo \cite{ref64} investigated non-minimal
    matter geometry coupling in the $f(R,T)$ gravity by using HEL. \\

    The hybrid scale factor has a transition behavior from deceleration to acceleration. Capozziello et al. \cite{ref65} studied
    the cosmographic bounds on cosmological deceleration-acceleration transition red shift in $f(R)$ gravity. The author considered
    a Tailor expansion of $f(z)$ in term of $ a(t) = \frac{1}{1+z}$ which for Friedmann equations, comes in the range $z \leq 2.$
    Capozziello et al. \cite{ref66} also extracts constraints on the transition red shift $z_{tr}$ in the frame work of $f(T)$ gravity
    which becomes compatible with the constraints predicted by $\Lambda$ CDM model at the 1-$\sigma$ confidence level. Their \cite{ref66}
    values seems to be slightly smaller than theoretical expectation, i.e., $z_{tr} = 0.74$ according to \cite{ref67}. Recently,
    Farooq et al. \cite{ref68} compile updated list of $38$ measurements of Hubble parameter $H(z)$ between red shifts $0 \leq z \leq 2.36 $
    and used them to put constraints on model parameters of constant and time-varying DE cosmological models, both spatially flat and curved.\\

Now  we will determine the constants  $\alpha$ and $\beta$ on the basis of the latest observational findings due to
Planck \cite{ref17}. The values of the cosmological parameters at present are as follows.
$(\Omega_{m})_{0}$= 0.30 $(\Omega_{k})_{0}= \pm0.005$, $(\omega_{de})_{0}=-1$, $(\Omega_{de})_{0}=0.70\pm0.005$, $H_0=0.07$
Gyr$^{-1}$ and present age $t_{0}=13.72$ Gyr.  \\

Using these values in Eq. (\ref{eq11}), we get the present value of the
deceleration parameter as $q_0 \simeq 0.55$. From Eq. (\ref{eq14}), we get
following
\begin{equation}
\label{eq15}
\alpha (1+z) H H_z =  \alpha (q+1)  H^2 = (H-\beta)^2 = \frac{\alpha^2}{t^2},
\end{equation}
where we have used $\dot{z}= -(1+z)H$ and $H_z$ means differentiation w.r.t. z.
From Eq. (\ref{eq15}) and Planck's results, we get the value of constants
$\alpha$ and $\beta$ as $\beta = 0.0397474\sim 0.04, \alpha = 0.415066 \sim
0.415$.

\section{Physical Properties of the model}
\subsection{Hubble Constant $H$ }
The determination of the two physical quantities  $H_{0}$ and  $q$ play an important role to describe the evolution of the universe.
$H_{0}$ provides us the rate of expansion of the universe which in
turn helps in estimating the age of the universe, whereas the deceleration parameter $q$ describes the decelerating or accelerating
phases during the  evolution of the universe. From the last two decades,  many attempts \cite{ref68}$-$\cite{ref74} have been made to estimate the value
of the Hubble constant as  $H_{0} = 72 \pm 8 km s^{-1} Mpc^{-1},~~  69.7^{+4.9}_{-5.0} km  s^{-1} Mpc^{-1},~~ 71 \pm 2.5 km s^{-1}
Mpc^{-1},~~ 70.4^{+1.3}_{-1.4} km s^{-1} Mpc^{-1},~~ 73.8 \pm 2.4 km s^{-1} Mpc^{-1} and~~ 67 \pm 3.2 km s^{-1} Mpc^{-1}$
respectively. For detail discussions readers are referred to Kumar and more latest Farook \cite{ref74,ref68}. \\

The exact solution of  Eq. (\ref{eq15}) is obtained for the Hubble constant $H$ as a function of redshift $z$ as follows

\begin{equation}
\label{eq16}
(H-\beta)^{\alpha}= A~\exp\left(\frac{\alpha\beta}{H-\beta}\right) (1+z),
\end{equation}
where the constant of integration A is obtained as $A = 0.134$ on the basis of the present value of $H (H_0=0.07$ Gy$^{-1})$.
A numerical solution of Eq. (\ref{eq16}) shows that the Hubble constant is an increasing function of red shift.
We present the following figures (1) and (2)  to illustrate the solution.
\begin{figure}[ht]
    \centering
    \includegraphics[width=10cm,height=5cm,angle=0]{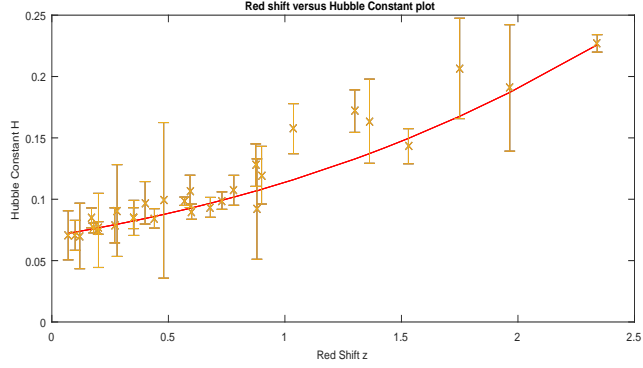}
    \caption{Plot of Hubble constant ($H$) versus redshift ($z$)}
\end{figure}
\begin{figure}[ht]
    \centering
    \includegraphics[width=10cm,height=5cm,angle=0]{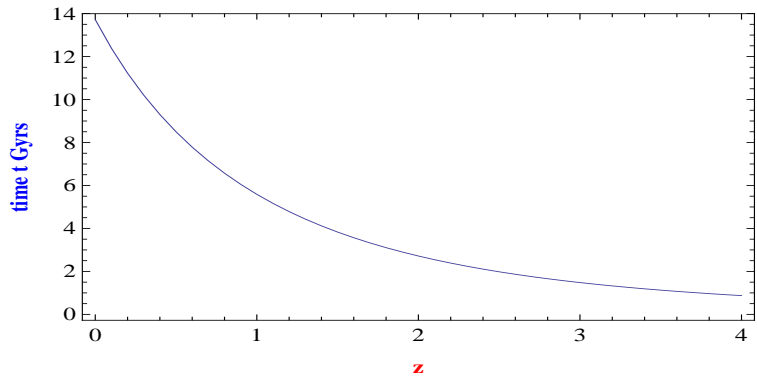}
    \caption{Variation of ($z$) versus  ($t$) }
\end{figure}

As  is clear from the figures, the Hubble constant   varies slowly over red shift and time. Various researchers \cite{ref75}$-$\cite{ref80}
 have estimated values of the Hubble constant at different red-shifts using a differential
age approach and galaxy clustering method. They have described various observed values of the Hubble constant $H_{ob}$ along
with corrections in the  range $0\leq z \leq 2$.  It is found that both observed and theoretical values tally considerably and support our model.

 In this figure $1$, cross signs are $31$ observed values of the Hubble constant $H_{ob}$ with corrections, whereas the linear curve is the theoretical graph of the Hubble constant $H$ as per our model. Figure $2$ is obtained from equation $\dot{z}= -(1+z)H(z)$. It plots the variation
of redshift $z$ with time $t$, which shows that in the early universe the redshift was more than at present. From this figure, we can convert redshift into time.
\subsection{ Transition from Deceleration to Acceleration }
Now we can obtain the deceleration parameter '$q$' in term of red shift '$z$' by using
Eqs. (\ref{eq15}) and (\ref{eq16}). We present the following figure (3) to illustrate the solution. This describes the
transition from deceleration to acceleration.

\begin{figure}[ht]
    \centering
    \includegraphics[width=10cm,height=5cm,angle=0]{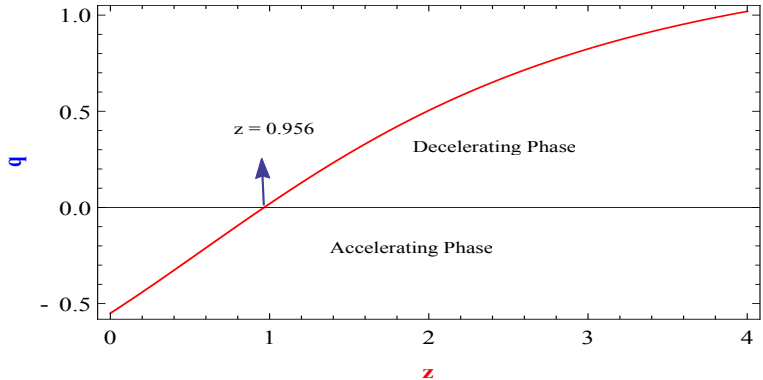}
    \caption{Variation of $q$ with $z$.}
\end{figure}

At $z=0.9557,~ \&~ 0.9558$, our model gives following values of Hubble constant $H$, deceleration  parameter $q$ and
and corresponding time.\\

$$H(0.9557)\to 0.111206,~ ~ q(0.9557)\to -0.0000124355, ~ ~t(0.9557)\to 5.81124 ,$$
and
$$H(0.9558)\to 0.111212, ~ ~ q(0.9558)\to 0.0000450098, ~ ~ t(0.9558)\to 5.81078 .$$

This means that the acceleration had begun at $ z\to 0.95575,t\to 5.81104~ Gyr, H\to 0.111209 ~Gyr^{-1} $.

\subsection{ DE Parameter  $\Omega_{de}$ and EoS  $\omega_{de}$ }

Now, from Eqs. (\ref{eq10}),(\ref{eq11}) and (\ref{eq16}), the density parameter $\Omega_{de}$ and EoS
parameter $\omega_{de}$ for DE are given by the following equations  and are solved numerically.
\begin{equation}
\label{eq17}
H^2 \Omega_{de} = H^2 - (\Omega_{m})_0 H^2_0 (1+z)^{3(1-\sigma)}
\end{equation}
and
\begin{equation}
\label{eq18}
\omega_{de}=\frac{(2-3\alpha)H^2-4\beta H+2 \beta^2}
{3 \alpha [H^2-H_0^2(\Omega_{m})_0(1+z)^{3(1-\sigma)}]},
\end{equation}
where we have taken  $(\Omega_{k})_0 = 0$ for the  present dust filled spatially
flat universe. We would take $ \sigma $ = 0.04 for numerical solutions to match
with latest observations. We solve Eqs. (\ref{eq17}) and (\ref{eq18}) with the help
of Eq. (\ref{eq16})  and present  following figures $3$ and $4$  to illustrate the
solution.
\begin{figure}[ht]
    \centering
    \includegraphics[width=10cm,height=5cm,angle=0]{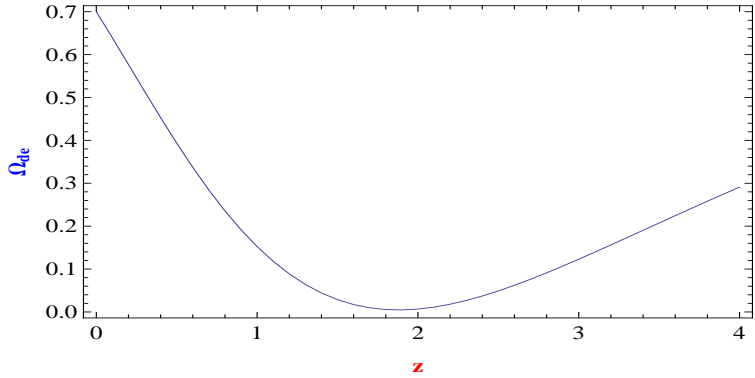}
    \caption{Plot of $\Omega_{de}$ versus redshift ($z$)}
\end{figure}
\begin{figure}[ht]
    \centering
    \includegraphics[width=10cm,height=5cm,angle=0]{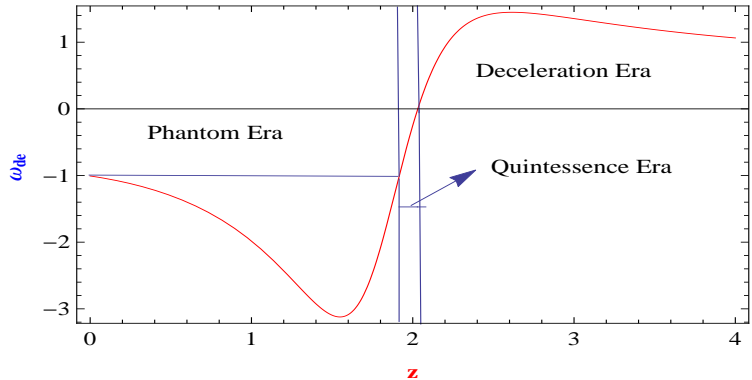}
    \caption{Plot of $\omega_{de}$ versus $z$. Phantom phase $(0 \leq z \leq 1.99)$, quintessence
        phase $ 1.99 \leq z \leq 2.0315$ and  deceleration phase  $z \geq 1.99$}
\end{figure}
Our model envisages that at present we are living in a phantom phase
$\omega_{(de)}\leq -1$.  In the past at $ z = 1.549~~ \omega_{(de)} = -3.12191$
was minimum, then it started increasing. This phase remains for the period $(0
\leq z \leq 1.99)$. Our universe entered into a quintessence phase at $ z =
1.99$, where $\omega_{de}$ comes up to $ -0.333123 $. As per our model, the
period for the quintessence phase is  the following
$$ 1.99 \leq z \leq 2.0315.$$ DE favors deceleration at $z \geq 1.99$.
The recent supernovae SNI $997ff$ at $z \simeq 1.7$ is consistent with a decelerated expansion at the epoch of high
emission \cite{ref72,ref79,ref80}.\\

As per our model, the present value of DE is 0.7. It decreases over the past, attains a  minimum value
$\Omega_{de}=0.0368568$ at $z= 1.834$, and then it again increases
with red shift. The dark energy density is approximately 29\% at red shift  4. Since dark
energy density is significant at this red shift, it might have strong implications on
structure formation, but at $z = 4$, EoS parameter  $\omega_{de}= 1.0532 $ is
positive, so it will favor deceleration and hence structure formation.

\subsection{Luminosity Distance}
The redshift-luminosity distance relation \cite{ref81} ia an important observational
tool to study the evolution of the universe. The expression for the luminosity
distance ($D_L$) is obtained in term of red-shift as the light coming out of a
distant luminous body gets red shifted due to the expansion of the universe. We
determine the flux of a source with the help of luminosity distance. It is given as
\begin{equation}\label{eq19}
D_{L}=a_{0} r (1+z),
\end{equation}
where r is the radial co ordinate of the source. In \cite{ref18}, $D_L$ is obtained
as
\begin{equation}
\label{eq20}
D_{L}=\frac{c(1+z)}{H_{0}}\int^z_0\frac{dz}{h(z)},~ h(z) = \frac{H}{H_0}
\end{equation}

\subsection{ Distance modulus $\mu$ and Apparent Magnitude $m_{b}$}

The distance modulus $\mu$  is  derived as  \cite{ref18}
\begin{eqnarray*}
    \mu & = &  m_{b}-M     \\
    & = &  5log_{10}\left(\frac{D_L}{Mpc}\right)+25 \\
    & = & 25+  5log_{10}\left[\frac{c(1+z)}{H_0} \int^z_0\frac{dz}{h(z)}\right].
\end{eqnarray*}
\begin{equation}\label{eq21}
\end{equation}

The absolute magnitude $M$ of a supernova \cite{ref18} is $ M=16.08-25+5log_{10}(H_{0}/.026c)$,
so we get following expression for the  apparent magnitude $m_b$

\begin{equation}\label{eq22}
m_{b}=16.08+ 5log_{10}\left[\frac{1+z}{.026} \int^z_0\frac{dz}{h(z)}\right].
\end{equation}

We solve Eqs. (\ref{eq20}), (\ref{eq21}) and (\ref{eq22}) with the help of Eq. (\ref{eq15}). Our theoretical results have been compared with  SNe Ia related
 $581$ data's from Pantheon compilation \cite{ref82} with possible error in the range ($0\leq z \leq1.4$) and
the derived model was found to be in good agreement with current observational constraints.
The following Figures $6$ \& $7$ depict the closeness of observational and theoretical results, thereby justifying
our model.
\begin{figure}[ht]
    \includegraphics[width=10cm,height=5cm,angle=0]{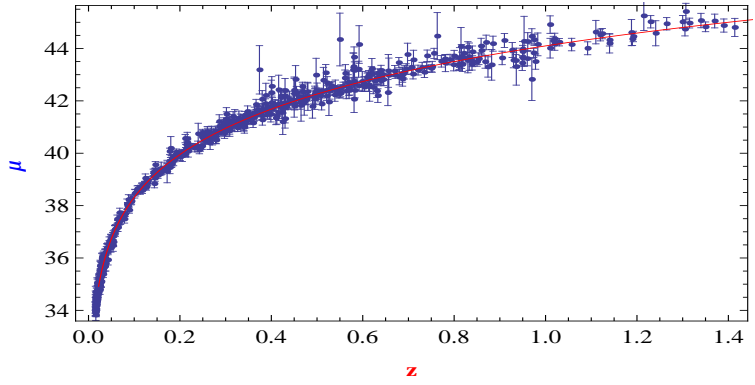}
\caption{Plot of distance modulus ($\mu$) versus red-shift ($z$). Crosses are  SNe Ia related
        $581$ data's from Pantheon compilation  with possible error.}
    \end{figure}

\begin{figure}[!ht]
    \includegraphics[width=10cm,height=5cm,angle=0]{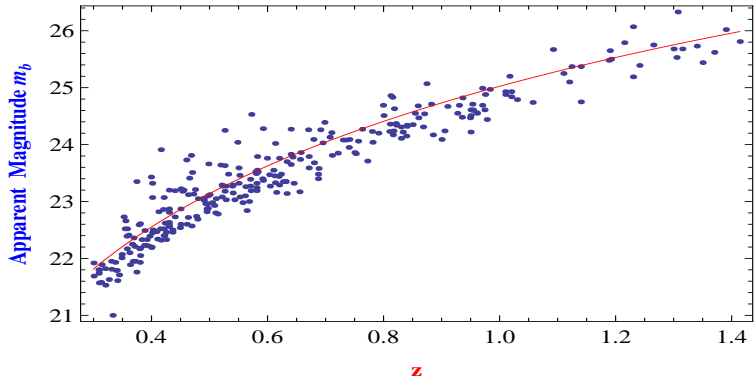}
    \caption{Plot of apparent magnitude ($m_b$) versus red-shift ($z$), Dots are  SNe Ia related
         $287$ data's from Pantheon compilation.  }
\end{figure}
\section{ Conclusions}
In this work, efforts were made to develop a cosmological model which satisfies the cosmological
principle and incorporates the latest developments which envisaged that our universe is accelerating due to DE.
We have also proposed a variable equation of state for DE in our model. We studied a model with  dust
and dark energy which shows a transition from deceleration to acceleration. We have successfully subjected our model to
various observational tests. The main findings of our model are itemized point-wise as follows.
\begin{itemize}
    \item The expansion of the universe is governed by a hybrid expansion law $ a(t)=t^{\alpha} \exp(\beta t ) $, where
    $\alpha =  0.415, \beta =  0.04 $. This describes the transition from  deceleration to acceleration.
    \item  Our model is based on the latest observational findings due to the Planck results \cite{ref17}. The model agrees with present cosmological
    parameters.\\
    $(\Omega_{m})_{0}$= 0.30 $(\Omega_{k})_{0}= \pm0.005$, $(\omega_{de})_{0}=-1$, $(\Omega_{de})_{0}=0.70\pm0.005$, $H_0=0.07$
    Gy$^{-1}$, $q_0 = 0.055$  and present age $t_{0}=13.72$ Gy.
    \item Our model has a variable equation of states  $\omega_{de}$ for the DE density.  Our model envisage that at present we are living in the phantom phase $\omega_{(de)}\leq -1$.  In the past at $ z = 1.549~~ \omega_{(de)} = -3.12191$
    was minimum, then it started increasing. This phase remains for the period $(0 \leq z \leq 1.99)$. Our universe entered into a quintessence
    phase at $ z = 1.99$ where $\omega_{de}$ comes up to $ -0.333123 $. As per our model, the
    period for the quintessence phase is  the following
    $$ 1.99 \leq z \leq 2.0315 $$. DE favors deceleration at $z \geq 1.99$.
    \item As per our model, the present value of DE is 0.7. It decreases over the past, attains a minimum value $\Omega_{de}=0.0368568$ at $z= 1.834$, and then it again increases
    with red shift.

    \item We have calculated the time at which acceleration had begun. The acceleration had begun at
    $ z\to 0.95575,t\to 5.81104~ Gyr, H\to 0.111209 ~Gyr^{-1} $. At this time $ \Omega_{de}=0.220369 $ and $\omega_{de}=-1.54715$
\end{itemize}

\section*{Acknowledgement} The authors (G. K. Goswami \& A. Pradhan) sincerely acknowledge the Inter-University Centre for Astronomy
and Astrophysics (IUCAA), Pune, India for providing facilities where part of this work

    \end{document}